\documentclass[11pt]{article}
\usepackage{amsmath}
\usepackage{amsfonts}
\usepackage{amssymb}
\usepackage{graphicx}

\def\bea{\begin{eqnarray}}
\def\eea{\end{eqnarray}}

\begin{document}
\begin{center}
\LARGE { \bf Correlation functions of BCFT
  }
\end{center}
\begin{center}
{\bf M. R. Setare\footnote{rezakord@ipm.ir} \\  V. Kamali\footnote{vkamali1362@gmail.com}}\\
 {\ Department of Science, Payame Noor University, Bijar, Iran}
 \\
 \end{center}
\vskip 3cm

\begin{abstract}
Boundary conformal field theory (BCFT) is the study of conformal
field theory (CFT) on manifolds with a boundary. We can use
conformal symmetry to constrain correlation functions of conformal
invariant fields. We compute two-point and three-point functions of conformal
invariant fields which live in semi-infinite space. For a
situation with a boundary condition in surface  $z=\overline{z}$
($t=0$), the results  agree  with gravity dual results. We also
explore representations of conformal group in two dimensions.
\\

\end{abstract}

\newpage

\section{Introduction}
The AdS/CFT correspondence \cite{ma,gu,wi}, which enables us to
study conformal field theory and non-perturbative quantum gravity
at the same time,  has been considered in during the past decade
. Holographic dual of a conformal field theory defined in domain
with a boundary was proposed in \cite{a}.  The main idea of
AdS/BCFT correspondence was started with asymptotically AdS
geometry with Neumann boundary condition on the metric as one
approaches to the boundary \cite{a}. The action of this theory is
given by Einstein-Hilbert action with a negative cosmological
constant which is added by boundary term \cite{a,mi,b}.
   \begin{eqnarray}\label{a}
   S=\frac{1}{16\pi G}\int_{M}d^3x\sqrt{-g}(R-2\Lambda)+\frac{1}{8\pi G}\int_{\partial M} d^2x\sqrt{-h}K+S_M
  \end{eqnarray}
 where $g$ and $h$ are the $3D$ bulk and $2D$ boundary metrics. The second term is the Gibbons-Hawking boundary term \cite{c}, which is given by $K=h^{ab}K_{ab}$. Extrinsic curvature $K_{ab}$  defined by $K_{ab}=\nabla_a n_b$ where $n$ is the unit vector normal to $\partial M$.  $S_M$ is action of some matter fields on the boundary $\partial M$. The geometry is modified by imposing two different boundary conditions on the metric. By this method the boundary is divided  into two parts $\partial M= N \bigcup Q$ where $\partial Q=\partial N$ \cite{a}. The metric has Neumann boundary condition on $Q$ and Dirichlet boundary condition on $N$. The variation of the action (\ref{a}) with respect to boundary metric $h_{ab}$ leads to
 \begin{eqnarray}\label{b}
 \delta S=\frac{1}{16\pi G}\int_{Q}\sqrt{-h}(K_{ab}\delta h^{ab}-Kh_{ab}\delta h^{ab}-8\pi G T_{ab}\delta h^{ab})d^2x
  \end{eqnarray}
  where
  \begin{eqnarray}\label{c}
  T_{ab}=-\frac{2}{\sqrt{-h}}\frac{\delta S_M}{\delta h_{ab}}
  \end{eqnarray}
 Neumann boundary condition was imposed by setting coefficients of $\delta h_{ab}$ to zero, so this equation is obtained
 \begin{eqnarray}\label{d}
 K_{ab}-h_{ab}K=8\pi G T_{ab}
  \end{eqnarray}
  If the boundary matter Lagrangian is a constant, the action has the following form
  \begin{eqnarray}\label{e}
   S=\frac{1}{16\pi G}\int_{M}d^3x\sqrt{-g}(R-2\Lambda)+\frac{1}{8\pi G}\int_{\partial M} d^2x\sqrt{-h}(K-T)+S_M
  \end{eqnarray}
 $T$ is interpreted as the tension of the boundary
surface Q. With this boundary condition ( Neumann boundary
condition on Q and Dirichlet boundary condition on N) the AdS
geometry is divided into two parts and the gravitational theory
lives in one part of this space. This modified geometry could
provide a holographic dual for BCFT \cite{a}. Boundary conformal
field theory (BCFT) defined in domains  with a boundary \cite
{ca}.  When CFT lives in semi-infinite  space, One sector of
conformal group is removed. For example, if we have a boundary
condition on surface $z=\overline{z}~( x+t=x-t)$ or $t=0$,
time-translation, Boost and time-spacial conformal transformation
are removed. So, two-point and three-point functions in this situation is
completely different from situation without boundary condition
(free space). In this paper we study correlation functions of
BCFT. By using some methods in conformal field theory \cite{ba},
we calculated two-point and three-point  functions in semi-infinite space and our
result for two-point function  agrees  with quantum gravity result \cite{b}. The paper
is organized as follow. In section (2) we build the
representations  of infinite conformal algebra in two dimensions
\cite{ba}. In section (3) we calculate two-point function in free
space. In section (4) we extended this calculation to the
space with boundary conditions. Then in section (5) we calculate  three-point function in free space and the space with boundary conditions. Finally in section (6), we close
by some concluding remarks.

\section{Representations of 2D conformal algebra  }
In this work  we  calculated correlation function of conformal
invariant fields which live in semi-infinite two-dimensional
space-time. Firstly, in this section we  build the
representations of infinite conformal algebra in two dimensions.
Conformal algebra is constructed by two copies  of non-centrally
Virasoro (Witt) algebra \cite{di}

\begin{eqnarray}\label{1}
[L_n,L_m]=(n-m)L_{n+m}~~~~~~~~~~~~~~~~~~~~~~~~~~~~~~~~~~~~~~~\\
\nonumber
[\overline{L}_n,\overline{L}_m]=(n-m)\overline{L}_{n+m}~~~~~~~~~~~~~~~~~~~~~~~~~~~~~~~~~~~~~~~\\
\nonumber
[L_n,\overline{L}_m]=0~~~~~~~~~~~~~~~~~~~~~~~~~~~~~~~~~~~~~~~~~~~~~~~~~~~~~~~
   \end{eqnarray}
where $L_{-1}$~($\overline{L}_{-1}$), $L_{0}$~($\overline{L}_0$)
and $L_{1}$~($\overline{L}_{1}$) are   translation, dilatation
and spacial conformal transformation (SCT) generators   in
$z$~($\overline{z}$) direction respectively. The representation
of conformal algebra is built  by considering operators which are
labeled by dilatations  eigenvalues. Local operators are given by
\begin{eqnarray}\label{2}
\mathcal{O}(z,\overline{z})=U\mathcal{O}(0)U^{-1}~~~~~where~~~~~~~U=e^{zL_{-1}+\overline{z}\overline{L}_{-1}}
  \end{eqnarray}

From Eq.(\ref{1}) we note that $[L_0,\overline{L}_0]=0$ and
$[L_0,L_n]\sim L_n$~($[\overline{L}_0,\overline{L}_n]\sim
\overline{L}_n$), so the representations of conformal algebra
should be labeled by eigenvalues of these two operators ($L_0$
and $\overline{L}_0$). We introduce local operators which are
simultaneous eigenstates of $L_0$ and $\overline{L}_0$.

\begin{eqnarray}\label{3}
[L_0,\mathcal{O}]=h\mathcal{O}~~~~~~~~~~~~~~~~~~~~[\overline{L}_0,\mathcal{O}]=\overline{h}\mathcal{O}
  \end{eqnarray}
  where $h$ ( $\overline{h}$) is left (right) conformal weight  ($h=\frac{\Delta+s}{2}, \overline{h}=\frac{\Delta-s}{2}$ where $\Delta$ is scaling dimension and $s$ is spin) \cite{di}.
In the following, the irreducible  representations of infinite
conformal algebra are considered. We use the Jacobi identity

\begin{eqnarray}\label{4}
[L_0,[L_n,\mathcal{O}]]=-[\mathcal{O},[L_0,L_n]]-[L_n,[\mathcal{O},L_0]]=n[\mathcal{O},L_n]+h[L_n,\mathcal{O}]\\
\nonumber
=(h-n)[L_n,\mathcal{O}]~~~~~~~~~~~~~~~~~~~~~~~~~~~~~~~~~~~~~~~~~~~~~~~~\\
\nonumber [\overline{L}_0,[\overline{L}_n,
\mathcal{O}]]=-[\mathcal{O},[\overline{L}_0,\overline{L}_n]-[\overline{L}_n,[\mathcal{O},\overline{L}_0]]=n[\mathcal{O},\overline{L}_n]+\overline{h}[\overline{L}_n,\mathcal{O}]\\
\nonumber
=(\overline{h}-n)[L_n,\mathcal{O}]~~~~~~~~~~~~~~~~~~~~~~~~~~~~~~~~~~~~~~~~~~~~~~~
  \end{eqnarray}
The above relations show that $L_n$~($\overline{L}_n$) thus lower
the value of left (right) conformal weight $h$~($\overline{h}$)
while $L_{-n}$ ($\overline{L}_{-n}$) raise it ($n>0$). We demand
that conformal weights is bounded from below, the primary
operators are  defined by these properties

\begin{eqnarray}\label{5}
[L_n,\mathcal{O}_p]=0~~~~~~~~~~~~~~~~~~~~~~[\overline{L}_n,\mathcal{O}_p]=0~~~~~~~~~~~~~~n>0
 \end{eqnarray}
By starting with a primary operators $\mathcal{O}_p$ and using the
relation (\ref{4}), we can build a tower of operators. These
operators form an irreducible representation of the conformal
algebra.
\section{Two-point function in free space}
We now turn to derive the consequences of conformal invariance
for the correlations. In general, we expect a quasi-primary
field  $\mathcal{O}$   to be characterized by its conformal
weights $h$ and $\overline{h}$ (These fields are invariant under
finite sub-group that is generated by sub-algebra
\{$L_{-1},\overline{L}_{-1}, L_0$,$\overline{L}_{0},
L_1,\overline{L}_1$\}). We would like to find the form of
two-point functions of the conformal invariant operators.
Firstly, we  find the form of the commutators $[L_n,
\mathcal{O}]$ and $[\overline{L}_n, \mathcal{O}]$

\begin{eqnarray}\label{6}
[L_n,\mathcal{O}(z,\overline{z})]=[L_n,U\mathcal{O}(0)U^{-1}]=[L_n,U]\mathcal{O}(0)U^{-1}+U\mathcal{O}(0)[L_n,U^{-1}]\\
\nonumber+U[L_n,\mathcal{O}(0)]U^{-1}=
U\{U^{-1}L_nU-L_n\}\mathcal{O}(0)U^{-1}~~~~~~~~~~~~~~~~~~~~~~~~~\\
\nonumber +U\mathcal{O}(0)\{L_n-U^{-1}L_n
U\}U^{-1}+\delta_{n,0}h\mathcal{O}(z,\overline{z})~~~~~~~~~~~~~~~~~~~~~~~~
  \end{eqnarray}
$U$ is defined in Eq.(\ref{2}). By using the Hausdorff formula we
get

\begin{eqnarray}\label{7}
U^{-1}L_nU=e^{-zL_{-1}-\overline{z}\overline{L}_{-1}}L_n
e^{zL_{-1}+\overline{z}\overline{L}_{-1}}=e^{-zL_{-1}}L_n
e^{zL_{-1}}
\\
\nonumber
=L_n+[L_n,zL_{-1}]+\frac{1}{2!}[[L_n,zL_{-1}],zL_{-1}]+...\\
\nonumber= \sum_{k=0}^{n+1}\frac{(n+1)!}{(n+1-k)!k!}(z)^k
L_{n-k}~~~~~~~~~~~~~~~~~~~~
  \end{eqnarray}
and
\begin{eqnarray}\label{8}
L_{n}^{'}=U^{-1}L_nU-L_n=\sum_{k=1}^{n+1}\frac{(n+1)!}{(n+1-k)!k!}(z)^k
L_{n-k}
  \end{eqnarray}
where $z=x+t$ and $\overline{z}=x-t$. From above relations, the
Eq.(\ref{6}) gives us
  \begin{eqnarray}
  [L_n,\mathcal{O}(z,\overline{z})]=U\{[L_n^{'},\mathcal{O}(0)]+\delta_{n,0}h\mathcal{O}(0)\}U^{-1}~~~~~~~~~\\
  \nonumber
  =z^{n+1}[L_{-1},\mathcal{O}(z,\overline{z})]+z^n(n+1)U[L_0,\mathcal{O}(0)]U^{-1}~~~~~~
  \end{eqnarray}
 Now we have $[L_{-1},\mathcal{O}]=\partial_{z}\mathcal{O}$ ($L_{-1}$ generates $z$-translation). Hence we obtain (for $n\geq-1$)
\begin{eqnarray}\label{9}
 [L_n,\mathcal{O}(z,\overline{z})]=(z^{n+1}\partial_{z}+(n+1)hz^n)\mathcal{O}
  \end{eqnarray}

We can exchange $L_n$ with  $\overline{L}_n$ and  using the above
steps (\ref{6}-\ref{9}). We get

  \begin{eqnarray}\label{10}
    [\overline{L}_n,\mathcal{O}(z,\overline{z})]=(\overline{z}^{n+1}\partial_{\overline{z}}+(n+1)h\overline{z}^n)\mathcal{O}
  \end{eqnarray}
From above equations (\ref{9} and \ref{10}),  we can constrain
correlation operators. We begin by considering two quasi-primary
operators $\mathcal{O}_1(z_1,\overline{z}_1)$ and
$\mathcal{O}_2(z_2,\overline{z}_2)$ which have conformal weights
($h_1,\overline{h}_1$) and ($h_2,\overline{h_2}$) respectively.\\
Two-point correlation function is defined as

\begin{eqnarray}\label{11}
  G(z_1,z_2;\overline{z}_1,\overline{z}_2)=<0\mid\mathcal{O}(z_1,\overline{z}_1)\mathcal{O}(z_2,\overline{z}_2)\mid0>
  \end{eqnarray}

In this section, we calculated two-point function in the bulk
without boundary condition. We are dealing with quasi-primary
fields, so we get four equations which would constrain the form
of the correlation function. Invariance under  $z$ and
$\overline{z}$ translation implies

  \begin{eqnarray}\label{12}
  <0\mid[L_{-1},G]\mid0>=0~~~~~~~~~~~~~~~~~<0\mid[\overline{L}_{-1},G]\mid0>=0\\
  \nonumber
  \Rightarrow
  \sum_{i=1}^{2}\partial_{z_i}G=0~~~~~~~~~~~~~~~~~~~~~~~~~~~\sum_{i=1}^{i=2}\partial_{\overline{z}_i}=0
  ~~~~~~~~~~~~~~\\
  \nonumber
  \Rightarrow
  G(z_1,z_2;\overline{z}_1,\overline{z}_2)=G(z,\overline{z})~~~~~~~~~z=z_1-z_2~~~~\overline{z}=\overline{z}_1-\overline{z}_2
  \end{eqnarray}
Invariance under dilatation of $z$ coordinate that generated by
$L_0$ requires

  \begin{eqnarray}\label{13}
  <0\mid[L_0,G(z,\overline{z})]\mid0>=0~~~~~~~~~~~~~~~~~~~~~~~~~~~~~~~~~~~~~~~~~~~~\\
  \nonumber
  \Rightarrow\sum_{i=1}^{2}(z_i\partial_{z_i}+h_i)G=0\Rightarrow(z_1\partial_{z_1}+z_2\partial_{z_2}+h_1+h_2)G=0~\\
  \nonumber
  \Rightarrow (z\partial_{z}+h)G=0 \Rightarrow
  G(z,\overline{z})=C(\overline{z})z^{-h}~~~~~~~~~h=h_1+h_2
  \end{eqnarray}

Similarly, invariance under $\overline{z}$-dilatation requires

  \begin{eqnarray}\label{14}
  <0\mid[\overline{L}_0,G(z,\overline{z})]\mid0>=0~~~~~~~~~~~~~~~~~~~~~~~~~~~~~~~~~~~~~~~~\\
  \nonumber
  \Rightarrow\sum_{i=1}^{2}(\overline{z}_i\partial_{\overline{z}_i}+\overline{h}_i)G=0\Rightarrow(\overline{z}_1\partial_{\overline{z}_1}+\overline{z}_2\partial_{\overline{z}_2}+\overline{h}_1+\overline{h}_2)G=0\\
  \nonumber
  \Rightarrow (z\partial_{z}+h)G=0 \Rightarrow
  G(z,\overline{z})=\overline{z}^{-\overline{h}}z^{-h}~~~~~~~~~\overline{h}=\overline{h}_1+\overline{h}_2
  \end{eqnarray}
New information come from requiring  SCT invariance which is
generated by $L_1$
\begin{eqnarray}\label{15}
<0\mid [L_1,G]\mid0>=0~~~~~~~~~~~~~~~~~~~~~~~~~~~~~~~~~~~~~~~\\
\nonumber \Rightarrow\sum_{i=1}^{2}(z_i^{2}\partial_{z_i}+2h_iz_i)G=0~~~~~~~~~~~~~~~~~~~~~~~~~~~~~~~~\\
\nonumber\Rightarrow
((z_1+z_2)z\partial_{z}+2h_1z_1+2h_2z_2)=0~~~~~~~~~~~~~~~~\\
\nonumber -(h_1+h_2)(z_1+z_2)+2h_1z_1+2h_2z_2=0\Rightarrow h_1=h_2
  \end{eqnarray}
and $\overline{L}_1$
  \begin{eqnarray}\label{16}
<0\mid
[\overline{L}_1,G]\mid0>=0~~~~~\Rightarrow~~~~~~~~~~~~~~~~\overline{h}_1=\overline{h}_2
  \end{eqnarray}

The final result is
  \begin{eqnarray}\label{17}
  G(z_1,z_2;\overline{z}_1,\overline{z}_2)=\delta_{h_1,h_2}\delta_{\overline{h}_1,\overline{h}_2}(z_1-z_2)^{-(h_1+h_2)}(\overline{z}_1-\overline{z}_2)^{-(\overline{h}_1+\overline{h}_2)}
  \end{eqnarray}
The above result is two-point function of two conformal invariant
fields which live in free space \cite{di}.
\section{Two-point function in semi-infinite space}
In this section  we calculated two-point function for
semi-infinite space with different boundary conditions, 1.
Boundary condition is in surface $z=0$ and boundary condition is
in surface $\overline{z}=0$.  2. The situation with a boundary on
surface $z=\overline{z}$. These three situations  are obviously
different, so we will receive  to different results.
\subsection{Two-point function in space with boundary condition in surface  $\overline{z}=0$ ($z=0$)}
We have a boundary condition in one dimension ($\overline{z}=0$),
so $\overline{z}$-translation and $\overline{z}$-SCT ,which are
generated by $\overline{L}_{-1}$ and $\overline{L}_1$
respectively, are obviously  removed. The reminded symmetry group
for this situation is
generated by subalgebra $[L_{-1},L_{1},L_{0},\overline{L}_{0}]$. (see for example \cite{henkel} and \cite{setare}).\\
Translation symmetry in $z$ coordinate, gives us

  \begin{eqnarray}\label{18}
  <0\mid[L_{-1},G]\mid0>=0 \Rightarrow
  G(z_1,z_2;\overline{z}_1,\overline{z}_2)=G(z;\overline{z}_1,\overline{z}_2)
  \end{eqnarray}
Invariance under $z$-dilatation implies

  \begin{eqnarray}\label{19}
  <0\mid[L_0,G(z,\overline{z}_1,\overline{z}_2)]\mid0>=0
  \Rightarrow
  G(z,\overline{z}_1,\overline{z}_2)=C_1(\overline{z}_1,\overline{z}_2)z^{-h}
  \end{eqnarray}

 and invariance under $\overline{z}$-dilatation implies
  \begin{eqnarray}\label{20}
  <0\mid[\overline{L}_0,G(z,\overline{z}_1,\overline{z}_2)]\mid0>=0~~~~~~~~~~~~~~~~~~~~~~~~~~~~~~~\\
  \nonumber\Rightarrow
  (\overline{z}_1\partial_{\overline{z}_1}+\overline{z}_2\partial_{\overline{z}_2}+\overline{h}_1+\overline{h}_2)C_1(\overline{z}_1,\overline{z}_2)=0~~~~~~~~~~~~~~~~~\\
  \nonumber
\Rightarrow
C_1(\overline{z}_1,\overline{z}_2)=\overline{z}_1^{-(\overline{h}_1+\overline{h}_2)}\Phi(\frac{\overline{z}_1}{\overline{z}_2})~~~~~~~~~~~~~~~~~~~~~~~~~~~~~~~
  \end{eqnarray}
where $\Phi$ is an arbitrary function. Finally, invariance under
$z$-SCT  gives us

  \begin{eqnarray}\label{21}
 <0\mid[\overline{L}_1,G(z,\overline{z}_1,\overline{z}_2)]\mid0>=0~~~~~\Rightarrow
 ~~~~~~h_1=h_2
  \end{eqnarray}

The form of two-point function in semi-infinite space with
boundary condition in $\overline{z}$ coordinate is given by

 \begin{eqnarray}\label{22}
 G(z_1,z_2;\overline{z}_1,\overline{z}_2)=\overline{z}_1^{-(\overline{h}_1+\overline{h}_2)}\Phi(\frac{\overline{z}_1}{\overline{z}_2})\delta_{h_1,h_2}(z_1-z_2)^{-(h_1+h_2)}
   \end{eqnarray}
We can exchange $z$ with  $\overline{z}$ and  using the above
steps (\ref{18}-\ref{22}). We get
\begin{eqnarray}\label{23}
G(z_1,z_2;\overline{z}_1,\overline{z}_2)=z_1^{-(h_1+h_2)}\Phi(\frac{z_1}{z_2})\delta_{\overline{h}_1,\overline{h}_2}(\overline{z}_1-\overline{z}_2)^{-(\overline{h}_1+\overline{h}_2)}
   \end{eqnarray}
The above relation is  two-point function in semi-infinite space
with a boundary condition in $z$ coordinate. We can see, these
results (\ref{22} and \ref{23}) are completely different from
result (\ref{17}).
\subsection{Two-point function in space with a boundary condition in surface $z=\overline{z}$}
 We have a boundary condition on surface $z=\overline{z}~( x+t=x-t)$ or $t=0$, so time-translation, Boost and time-SCT are removed. These transformations are generated by $L_n-\overline{L}_n$ ($n=-1,0,1$) generators. In this situation the reminded symmetry group is generated by one copy of non-centrally Virasoro algebra ($\mathfrak{L}_n=L_n+\overline{L}_{n}$) \cite{b}.
  \begin{eqnarray}\label{24}
  [\mathfrak{L}_n,\mathfrak{L}_m]=[L_n,L_m]+[\overline{L}_n,\overline{L}_m]=(n-m)(L_{n+m}+\overline{L}_{n+m})=(n-m)\mathfrak{L}_{n+m}
   \end{eqnarray}
From Eqs.(\ref{9},\ref{10}) the form of commutator
$[\mathfrak{L}_n,\mathcal{O}]$ is given by
    \begin{eqnarray}\label{26}
 [\mathfrak{L}_n,\mathcal{O}]=(z^{n+1}\partial_z+\overline{z}^{n+1}\partial_{\overline{z}}+(n+1)hz^n+(n+1)\overline{h}\overline{z}^n)\mathcal{O}
   \end{eqnarray}
   From above equation one can calculate two-point function of two quasiprimary operator $\mathcal{O}_1$ and $\mathcal{O}_2$ with conformal weights ($h_1$,$\overline{h}_1$) and ($h_2$,$\overline{h}_2$).
 Invariance under $x$-translation implies ($x=\frac{z+\overline{z}}{2}$)
 \begin{eqnarray}\label{27}
 <0\mid [\mathfrak{L}_{-1},G(z_1,z_2,\overline{z}_1,\overline{z}_2)]\mid0>=0\Rightarrow\sum_{i=1}^{2}(\partial_{z_i}+\partial_{\overline{z}_i})G=0
   \end{eqnarray}
 we make the ansatz
   \begin{eqnarray}\label{28}
   G(z_1,z_2,\overline{z}_1,\overline{z}_2)=G_1(|z_1-z_2|)+G_2(|\overline{z}_1-\overline{z}_2|)+G_3(|z_1-\overline{z}_2|)+G_4(|\overline{z}_1-z_2|)
   \end{eqnarray}
 Invariance under dilatation which is generated by $\mathfrak{L}_0$ requires
\begin{eqnarray}\label{29}
<0\mid[\mathfrak{L}_{0},G(z_1,z_2,\overline{z}_1,\overline{z}_2)]\mid0>=0~~~~~~~~~~~~~~~~~~~~~~~~~~~~\\
\nonumber \Rightarrow
\sum_{i=1}^2(z_i\partial_{z_i}+h_i+\overline{z}_i\partial_{\overline{z}_i}+\overline{h}_i)G=0~~~~~~~~~~~~~~~~~~~~~~~~~~
   \end{eqnarray}
   so
   \begin{eqnarray}\label{30}
   (z_1\partial_{z_1}+z_2\partial_{z_2}+\Delta_1+\Delta_2)G_1=0~~~~~~~~~~~~~~~~~~~~\\
   \nonumber
   (\overline{z}_1\partial_{\overline{z}_1}+\overline{z}_2\partial_{\overline{z}_2}+\Delta_1+\Delta_2)G_2=0~~~~~~~~~~~~~~~~~~~~\\
   \nonumber
   (z_1\partial_{z_1}+\overline{z}_2\partial_{\overline{z}_2}+\Delta_1+\Delta_2)G_3=0~~~~~~~~~~~~~~~~~~~~\\
   \nonumber
   (\overline{z}_1\partial_{\overline{z}_1}+z_2\partial_{z_2}+\Delta_1+\Delta_2)G_4=0~~~~~~~~~~~~~~~~~~~~\\
   \end{eqnarray}
where $\Delta_i=h_i+\overline{h}_i$. Final results for $G_i$ are
\begin{eqnarray}\label{30}
G_1\sim\frac{1}{\mid z_1-z_2\mid^{\Delta_1+\Delta_2}}~~~~~~~~~~~~~~~~~~~~~G_2\sim\frac{1}{\mid \overline{z}_1-\overline{z}_2\mid^{\Delta_1+\Delta_2}}\\
\nonumber G_3\sim\frac{1}{\mid
z_1-\overline{z}_2\mid^{\Delta_1+\Delta_2}}~~~~~~~~~~~~~~~~~~~~~G_i\sim\frac{1}{\mid
\overline{z}_1-z_2\mid^{\Delta_1+\Delta_2}}
   \end{eqnarray}
spatial-SCT which generated by $\mathfrak{L}_1$ gives new
information.
\begin{eqnarray}\label{31}
<0\mid[\mathfrak{L}_{1},G(z_1,z_2,\overline{z}_1,\overline{z}_2)]\mid0>=0\Rightarrow~~\Delta_1=\Delta_2=\Delta
   \end{eqnarray}
Final result of Two-point function in this situation (boundary on
surface $z=\overline{z}$) is
\begin{eqnarray}\label{32}
G(z_1,z_2,\overline{z}_1,\overline{z}_2)=\frac{1}{4}(\frac{1}{\mid z_1-z_2\mid^{2\Delta}}\pm\frac{1}{\mid \overline{z}_1-\overline{z}_2\mid^{2\Delta}}\\
\nonumber \pm\frac{1}{\mid
z_1-\overline{z}_2\mid^{2\Delta}}+\frac{1}{\mid
\overline{z}_1-z_2\mid^{2\Delta}})
   \end{eqnarray}
where $\overline{z}_{1,2}$ are images of $z_{1,2}$ and, the plus
and minus signs correspond to Neumann and Dirichlet boundary
condutions, respectively. We note that the above result agrees
with correlation function of scalar operators which was
calculated by Alishahiha and Fareghbal from gravity dual of BCFT
\cite{b}. Now, we  introduce new variable

\begin{eqnarray}\label{}
\zeta=\frac{|z_1-z_2||\overline{z}_1-\overline{z}_2|}{|z_1-\overline{z}_1||z_2-\overline{z}_2|}
   \end{eqnarray}
Two-point function of BCFT may be written in the following form
\begin{eqnarray}\label{}
G=[\frac{|z_1-\overline{z}_1||z_2-\overline{z}_2|}{|z_1-z_2||\overline{z}_1-\overline{z}_2||z_1-\overline{z}_2||\overline{z}_1-z_2|}]^{\Delta}F(\zeta)
   \end{eqnarray}
where
\begin{eqnarray}\label{}
F(\zeta)=(constant)[(\zeta+1)^{\Delta}\pm\zeta^{\Delta}]
   \end{eqnarray}
This equation agrees with result \cite{ca}.

\section{Three-point function}
From above method, we will construct the three-point function
of three quasi-primary operators $\mathcal{O}_1$, $\mathcal{O}_2$
and $\mathcal{O}_3$ which have conformal weights
$(h_1,\overline{h}_1)$, $(h_2,\overline{h}_2)$ and
$(h_3,\overline{h}_3)$. First, we calculated three-point function
in free space without boundary condition. Three-point function is
defined as.
\begin{eqnarray}\label{t1}
G^{(3)}(z_1,z_2,z_3,\overline{z}_1,\overline{z}_2,\overline{z}_3)=<0\mid\mathcal{O}_1\mathcal{O}_2\mathcal{O}_3\mid0>=G_{z}^{(3)}(z_1,z_2,z_3)G_{\overline{z}}^{(3)}(\overline{z}_1,\overline{z}_2\overline{z}_3)
\end{eqnarray}
$G_{z}^{(3)}$ is constrained by $L_{-1}$, $L_{0}$ and $L_1$
generators ($G_{\overline{z}}^{(3)}$ is constrained by
$\overline{L}_{-1}$, $\overline{L}_{0}$ and $\overline{L}_1$
generators). We calculated $z$-sector of three-point function.
Invariance under $z$ translation implies.
\begin{eqnarray}\label{t2}
 <0\mid[L_{-1},G_z^{(3)}]\mid0>=0~~~~~~~~~~~~\Rightarrow \sum_{i=1}^{3}\partial_{z_i}G_{z}^{(3)}=0~~~~~~~~~~~~~~\\
 \nonumber
 \Rightarrow G_{z}^{(3)}(z_1,z_2,z_3)=G_{z}^{(3)}(z_a,z_b)~~~~~~~~~~z_a=z_1-z_3~~~~~z_b=z_2-z_3
\end{eqnarray}
Dilatation of $z$-coordinate which is generated by $L_0$
constrains Three-point function as
\begin{eqnarray}\label{t3}
<0\mid[L_0,G_z^{(3)}]\mid0>=0~~~~~~~~~~~~~~~~~~~~~~~~~~~~~~~~~~~~~~~~~~~~\\
  \nonumber
  \Rightarrow\sum_{i=1}^{2}(z_i\partial_{z_i}+h_i)G_z^{(3)}=0~~~~~~~~~~~~~~~~~~~~~~~~~~~~~~~~~~~~~~~~~~~~\\
  \nonumber
  \Rightarrow(z_1\partial_{z_1}+z_2\partial_{z_2}+z_3\partial_{z_3}+h)G_z^{(3)}=0~~~~~h=h_1+h_2+h_3\\
  \nonumber
  \Rightarrow (z_a\partial_{z_a}+z_b\partial_{z_b}+h)G_z^{(3)}=0~~~~~~~~~~~~~~~~~~~~~~~~~~~~~~~~~~~~~~
\end{eqnarray}
$SCT$-invariance in $z$-direction gives the following equation
\begin{eqnarray}\label{t4}
<0\mid[L_1,G_z^{(3)}]\mid0>=0~~~~~~~~~~~~~~~~~~~~~~~~~~~~~~~~~~~~~~~~~~~~~~~~~~~~~\\
\nonumber
\Rightarrow (z_1^2\partial_{z_1}+z_2^2\partial_{z_2}+z_3\partial_{z_3}+2z_1h_1+2z_2h_2+2z_3h_3)G_z^{(3)}=0~~~~~~\\
\nonumber
\Rightarrow ((z_1^2-z_3^2)\partial_{z_a}+(z_2^2-z_3^2)\partial_{z_b}+2z_1h_1+2z_2h_2+2z_3h_3)G_z^{(3)}=0\\
\nonumber \Rightarrow
(z_a^2\partial_{z_a}+z_b^2\partial_{z_b}+2z_ah_1+2z_bh_2)G_{z}^{(3)}=0~~~~~~~~~~~~~~~~~~~~~~~~~~~~
\end{eqnarray}
We make the following ansatz for the $z$-sector of three-point
function
\begin{eqnarray}\label{t5}
G_z^{(3)}=z_a^{\alpha}.z_b^{\beta}.(z_a-z_b)^{\gamma}
\end{eqnarray}
From Eqs.(\ref{t3}) and (\ref{t4}) we have
\begin{eqnarray}\label{t6}
\alpha+\beta+\gamma+h_1+h_2+h_3=0\\
\nonumber
\alpha+\gamma+2h_1=0~~~~~~~~~~~~~~~~~\\
\nonumber \beta+\gamma+2h_2=0~~~~~~~~~~~~~~~~~~
\end{eqnarray}
So, $z$-sector of three-point function is given by
\begin{eqnarray}\label{t7}
G_z^{(3)}=(z_1-z_2)^{(h_2-h_1-h_3)}(z_2-z_3)^{(h_1-h_2-h_3)}(z_1-z_2)^{(h_3-h_1-h_2)}
\end{eqnarray}
$G_{\overline{z}}^{(3)}$ is calculated by this method.
\begin{eqnarray}\label{t8}
G_{\overline{z}}^{(3)}=(\overline{z}_1-\overline{z}_2)^{(\overline{h}_2-\overline{h}_1-\overline{h}_3)}(\overline{z}_2-\overline{z}_3)^{(\overline{h}_1-\overline{h}_2-\overline{h}_3)}(\overline{z}_1-\overline{z}_2)^{(\overline{h}_3-\overline{h}_1-\overline{h}_2)}
\end{eqnarray}
Final result of three-point function is given by
\begin{eqnarray}\label{t9}
G^{(3)}=(z_1-z_2)^{(h_2-h_1-h_3)}(z_2-z_3)^{(h_1-h_2-h_3)}(z_1-z_2)^{(h_3-h_1-h_2)}\\
\nonumber\times(\overline{z}_1-\overline{z}_2)^{(\overline{h}_2-\overline{h}_1-\overline{h}_3)}(\overline{z}_2-\overline{z}_3)^{(\overline{h}_1-\overline{h}_2-\overline{h}_3)}(\overline{z}_1-\overline{z}_2)^{(\overline{h}_3-\overline{h}_1-\overline{h}_2)}
\end{eqnarray}
In the following, the form of three-point function for three
quasi-primary field,  in semi-infinite space with boundary
condition in $\overline{z}=0$ is calculated. As we have seen the
symmetry group of this situation is generated by subalgebra
$[L_{-1}, L_0, L_1, \overline{L}_0]$. From invariance under
$[L_{-1}, L_0, L_1]$, $z$-sector of the three-point function is
unchange (\ref{t7}). Invariance under $\overline{z}$-sector gives
the following equation
\begin{eqnarray}\label{t10}
[\overline{L}_0,G_{\overline{z}}^{(3)}]=0~~~~~~~~~~~~~~~~~~~~~~~~~~~~~~~~~~~~~~~~~~~~~~~~~~~~~~~~~~~~~\\
\nonumber
\Rightarrow\sum_{i=1}^{2}(\overline{z}_i\partial_{\overline{z}_i}+\overline{h}_i)G_z^{(3)}=0~~~~~~~~~~~~~~~~~~~~~~~~~~~~~~~~~~~~~~~~~~~~\\
\nonumber
\Rightarrow(\overline{z}_1\partial_{\overline{z}_1}+\overline{z}_2\partial_{\overline{z}_2}+\overline{z}_3\partial_{\overline{z}_3}+\overline{h})G_{\overline{z}}^{(3)}=0~~~~~\overline{h}=\overline{h}_1+\overline{h}_2+\overline{h}_3\\
\end{eqnarray}
We make the following ansatz for this sector
\begin{eqnarray}\label{t11}
G_{\overline{z}}^{(3)}=\overline{z}_1^{-\overline{h}}\phi(\frac{\overline{z}_2}{\overline{z}_3})+\sum
(exchanging 1\rightarrow 2~~ and~~ 1\rightarrow 3 )
\end{eqnarray}
So, three-point function in semi-infinite space with boundary
condition in $\overline{z}=0$ has this form
\begin{eqnarray}\label{t12}
G^{(3)}=(z_1-z_2)^{(h_2-h_1-h_3)}(z_2-z_3)^{(h_1-h_2-h_3)}(z_1-z_2)^{(h_3-h_1-h_2)}\\
\nonumber
\times(\overline{z}_1^{-\overline{h}}\phi(\frac{\overline{z}_2}{\overline{z}_3})+\sum
(exchanging 1\rightarrow 2~~ and~~ 1\rightarrow 3 ))
\end{eqnarray}
We can exchange $z$ with  $\overline{z}$ and  using the above
steps. We get
\begin{eqnarray}\label{t13}
G^{(3)}=(\overline{z}_1-\overline{z}_2)^{(\overline{h}_2-\overline{h}_1-\overline{h}_3)}(\overline{z}_2-\overline{z}_3)^{(\overline{h}_1-\overline{h}_2-\overline{h}_3)}(\overline{z}_1-\overline{z}_2)^{(\overline{h}_3-\overline{h}_1-\overline{h}_2)}\\
\nonumber \times(z_1^{-h}\phi(\frac{z_2}{z_3})+\sum (exchanging
1\rightarrow 2~~ and~~ 1\rightarrow 3 ))
\end{eqnarray}
The above relation is three-point function in semi-infinite space
with a boundary condition in $z$ coordinate. We can see, these
results ((\ref{t12}) and (\ref{t13})) are completely different
from result (\ref{t9}). The situation with another boundary
condition is considered, we have a boundary condition in  surface
$z=\overline{z}$ or $t=0$. Time translation, Boost invariance and
time component of SCT is removed. The reminded symmetry-subgroup
is generated by $L_n+\overline{L}_n$ ($n=-1,0,1$) generators. We
constrained  Three-point function by this subgroup. First,
invariance under translation which is generated by
$L_{-1}+\overline{L}_{-1}$
\begin{eqnarray}\label{t14}
<0\mid[L_{-1}+\overline{L}_{-1},G]\mid0>=0~~~~~~~~~~~~~~~~~~~~~~~~~~~\\
\nonumber \Rightarrow
\sum_{i=1}^{3}(\partial_{z_i}+\partial_{\overline{z}_i})G=0~~~~~~~~~~~~~~~~~~~~~~~~~~~~~~~~~~~~~\\
\nonumber
(\partial_{z_1}+\partial_{\overline{z}_1}+\partial_{z_2}+\partial_{\overline{z}_2}+\partial_{z_3}+\partial_{\overline{z}_3})G=0~~~~~~~~~~~~~
\end{eqnarray}
From this equation, we make the ansatz
\begin{eqnarray}\label{t15}
G=a_1G_1(\mid z_1-z_2\mid,\mid z_1-z_3\mid,\mid
z_2-z_3\mid)+a_2G_2(\mid z_1-\overline{z}_2\mid,\mid
z_1-z_3\mid,\mid z_2-\overline{z}_3\mid)\\
\nonumber +a_3G_3(\mid z_1-z_2 \mid,\mid
z_1-\overline{z}_3\mid,\mid z_2-\overline{z}_3\mid)+a_4G_4(\mid
z_1-\overline{z}_3\mid,\mid z_1-\overline{z}_3
\mid,\mid\overline{z}_2-\overline{z}_3\mid)\\
\nonumber +a_5G_5(\mid \overline{z}_1-\overline{z}_2\mid,\mid
\overline{z}_1-\overline{z}_3 \mid,\mid
\overline{z}_2-\overline{z}_3\mid)+a_6G(\mid \overline{z}_1-z_2
\mid,\mid z_2-\overline{z}_3\mid,\mid
\overline{z}_1-\overline{z}_3\mid)\\
\nonumber a_7G_7(\mid
\overline{z}_1-\overline{z}_2\mid,\mid\overline{z}_1-z_3\mid,\mid\overline{z}_2-z_3\mid)+a_8G_8(\mid
\overline{z}_1-z_2 \mid,\mid\overline{z}_1-z_3\mid,\mid
z_2-z_3\mid)
\end{eqnarray}
where $a_i$ ($i=1...8$) are arbitrary constants. Dilatation
constrains three-point function as
\begin{eqnarray}\label{t16}
<0\mid[L_0+\overline{L}_0,
G]\mid0>=0~~~~~~~~~~~~~~~~~~~~~~~~~~~\\
\nonumber
(z_1\partial_{z_1}+z_2\partial_{z_2}+z_3\partial_{z_3}+\Delta_1+\Delta_2+\Delta_3)G_1=0\\
\nonumber
(z_1\partial_{z_1}+\overline{z}_2\partial_{\overline{z}_2}+\overline{z}_3\partial_{\overline{z}_3}+\Delta_1+\Delta_2+\Delta_3)G_2=0\\
\nonumber
(z_1\partial_{z_1}+z_2\partial_{z_2}+z_3\partial_{z_3}+\Delta_1+\Delta_2+\Delta_3)G_3=0\\
\nonumber
(z_1\partial_{z_1}+\overline{z}_2\partial_{\overline{z}_2}+\overline{z}_3\partial_{\overline{z}_3}+\Delta_1+\Delta_2+\Delta_3)G_4=0\\
\nonumber
(\overline{z}_1\partial_{\overline{z}_1}+\overline{z}_2\partial_{\overline{z}_2}+\overline{z}_3\partial_{\overline{z}_3}+\Delta_1+\Delta_2+\Delta_3)G_5=0\\
\nonumber
(\overline{z}_1\partial_{\overline{z}_1}+z_2\partial_{z_2}+\overline{z}_3\partial_{\overline{z}_3}+\Delta_1+\Delta_2+\Delta_3)G_6=0\\
\nonumber
(\overline{z}_1\partial_{\overline{z}_1}+\overline{z}_2\partial_{\overline{z}_2}+z_3\partial_{z_3}+\Delta_1+\Delta_2+\Delta_3)G_7=0\\
\nonumber
(\overline{z}_1\partial_{\overline{z}_1}+z_2\partial_{z_2}+z_3\partial_{z_3}+\Delta_1+\Delta_2+\Delta_3)G_8=0
\end{eqnarray}
Invariance under spatial-SCT gives
\begin{eqnarray}\label{t17}
(z_1^2\partial_{z_1}+z_2^2\partial_{z_2}+z_3^2\partial_{z_3}+2z_1\Delta_1+2z_2\Delta_2+2z_3\Delta_3)G_1=0\\
\nonumber
(z_1^2\partial_{z_1}+\overline{z}_2^2\partial_{\overline{z}_2}+z_3^2\partial_{z_3}+2z_1\Delta_1+2\overline{z}_2\Delta_2+2z_3\Delta_3)G_2=0\\
\nonumber
(z_1^2\partial_{z_1}+z_2^2\partial_{z_2}+\overline{z}_3^2\partial_{\overline{z}_3}+2z_1\Delta_1+2z_2\Delta_2+2\overline{z}_3\Delta_3)G_3=0\\
\nonumber
(z_1^2\partial_{z_1}+\overline{z}_2^2\partial_{\overline{z}_2}+\overline{z}_3^2\partial_{\overline{z}_3}+2z_1\Delta_1+2\overline{z}_2\Delta_2+2\overline{z}_3\Delta_3)G_4=0\\
\nonumber
(\overline{z}_1^2\partial_{\overline{z}_1}+\overline{z}_2^2\partial_{\overline{z}_2}+\overline{z}_3^2\partial_{\overline{z}_3}+2\overline{z}_1\Delta_1+2\overline{z}_2\Delta_2+2\overline{z}_3\Delta_3)G_5=0\\
\nonumber
(\overline{z}_1^2\partial_{\overline{z}_1}+z_2^2\partial_{z_2}+\overline{z}_3^2\partial_{\overline{z}_3}+2\overline{z}_1\Delta_1+2z_2\Delta_2+2\overline{z}_3\Delta_3)G_6=0\\
\nonumber
(\overline{z}_1^2\partial_{\overline{z}_1}+\overline{z}_2^2\partial_{\overline{z}_2}+z_3^2\partial_{z_3}+2\overline{z}_1\Delta_1+2\overline{z}_2\Delta_2+2z_3\Delta_3)G_7=0\\
\nonumber
(\overline{z}_1^2\partial_{\overline{z}_1}+z_2^2\partial_{z_2}+\overline{z}_3^2\partial_{\overline{z}_3}+2\overline{z}_1\Delta_1+2z_2\Delta_2+2\overline{z}_3\Delta_3)G_8=0\\
\nonumber
\end{eqnarray}
From Eqs. (\ref{t16}) and (\ref{t17}) and using steps
(\ref{t2})-(\ref{t7}) we calculated $G_i$
\begin{eqnarray}\label{t18}
G_1=(z_1-z_3)^{\Delta_2-\Delta_1-\Delta_3}(z_2-z_3)^{\Delta_1-\Delta_2-\Delta_3}(z_1-z_2)^{\Delta_3-\Delta_1-\Delta_2}\\
\nonumber
G_2=(z_1-z_3)^{\Delta_2-\Delta_1-\Delta_3}(\overline{z}_2-z_3)^{\Delta_1-\Delta_2-\Delta_3}(z_1-\overline{z}_2)^{\Delta_3-\Delta_1-\Delta_2}\\
\nonumber
G_3=(z_1-\overline{z}_3)^{\Delta_2-\Delta_1-\Delta_3}(z_2-\overline{z}_3)^{\Delta_1-\Delta_2-\Delta_3}(z_1-z_2)^{\Delta_3-\Delta_1-\Delta_2}\\
\nonumber
G_4=(z_1-\overline{z}_3)^{\Delta_2-\Delta_1-\Delta_3}(\overline{z}_2-\overline{z}_3)^{\Delta_1-\Delta_2-\Delta_3}(z_1-\overline{z}_2)^{\Delta_3-\Delta_1-\Delta_2}\\
\nonumber
G_5=(\overline{z}_1-\overline{z}_3)^{\Delta_2-\Delta_1-\Delta_3}(\overline{z}_2-\overline{z}_3)^{\Delta_1-\Delta_2-\Delta_3}(\overline{z}_1-\overline{z}_2)^{\Delta_3-\Delta_1-\Delta_2}\\
\nonumber
G_6=(\overline{z}_1-\overline{z}_3)^{\Delta_2-\Delta_1-\Delta_3}(z_2-\overline{z}_3)^{\Delta_1-\Delta_2-\Delta_3}(\overline{z}_1-z_2)^{\Delta_3-\Delta_1-\Delta_2}\\
\nonumber
G_7=(\overline{z}_1-z_3)^{\Delta_2-\Delta_1-\Delta_3}(\overline{z}_2-z_3)^{\Delta_1-\Delta_2-\Delta_3}(\overline{z}_1-\overline{z}_2)^{\Delta_3-\Delta_1-\Delta_2}\\
\nonumber
G_8=(\overline{z}_1-z_3)^{\Delta_2-\Delta_1-\Delta_3}(z_2-z_3)^{\Delta_1-\Delta_2-\Delta_3}(\overline{z}_1-z_2)^{\Delta_3-\Delta_1-\Delta_2}\\
\end{eqnarray}
Final result is given by
\begin{eqnarray}\label{t19}
G=\frac{1}{8}(G_1\pm G_2\pm G_3\pm G_4\pm G_5\pm G_6\pm G_7\pm
G_8)
\end{eqnarray}
The plus and minus signs correspond to Neumann and Dirichlet
boundary conditions, respectively.

\section{Conclusion}
We can use 2D conformal group to constrain correlation functions.
Two-point and three-point functions of conformal invariant fields in free space
were  found in \cite{di}. We computed these results by using some
methods in conformal field theory. Also by these methods we can
compute correlation functions  of conformal invariant fields
which live in semi-infinite spaces. In this paper we consider
four situations. 1. A system without any boundary condition. 2. A
system with a boundary condition on surface $z=0$. 3. A system
with a boundary condition on surface $\overline{z}=0$. 4. A
system in semi-infinite space with a boundary condition on
surface $z=\overline{z}$ or $t=0$. The main results of this work
are the explicit expressions for two-point functions of conformal
invariant fields in semi-infinite spaces as given in Eqs.
(\ref{22}), (\ref{23}) and (\ref{32}) and three-point function in semi-infinite space as given in Eqs. (\ref{t12}), (\ref{t13}) and (\ref{t19}). The result (\ref{32}) agrees with two-point function of scalar operators which was
calculated from gravity dual of BCFT \cite{b}.

\end{document}